**Title:** Directed Acyclic Graphs and causal thinking in clinical risk prediction modeling

**Authors:** Marco Piccininni, Stefan Konigorski, Jessica L Rohmann, Tobias Kurth

**Affiliations:**

Institute of Public Health, Charité - Universitätsmedizin Berlin, Berlin, Germany (Marco Piccininni, Jessica L Rohmann, Tobias Kurth); Digital Health & Machine Learning Research Group, Hasso Plattner Institute for Digital Engineering, Potsdam, Germany (Stefan Konigorski); Hasso Plattner Institute for Digital Health at Mount Sinai, Icahn School of Medicine at Mount Sinai, New York, USA (Stefan Konigorski)

**Corresponding author:**

Marco Piccininni
Institute of Public Health
Charité - Universitätsmedizin Berlin
Chariteplatz 1
10117 Berlin
marco.piccininni@charite.de

**Abstract**

**Background:** In epidemiology, causal inference and prediction modeling methodologies have been historically distinct. Directed Acyclic Graphs (DAGs) are used to model *a priori* causal assumptions and inform variable selection strategies for causal questions. Although tools originally designed for prediction are finding applications in causal inference, the counterpart has remained largely unexplored. The aim of this theoretical and simulation-based study is to assess the potential benefit of using DAGs in clinical risk prediction modeling.

**Methods and Findings:** We explore how incorporating knowledge about the underlying causal structure can provide insights about the transportability of diagnostic clinical risk prediction models to different settings. A single-predictor model in the causal direction is likely to have better transportability than one in the anticausal direction. We further probe whether causal knowledge can be used to improve predictor selection. We empirically show that the Markov Blanket, the set of variables including the parents, children, and parents of the children of the outcome node in a DAG, is the optimal set of predictors for that outcome.

**Conclusions:** Our findings challenge the generally accepted notion that a change in the distribution of the predictors does not affect diagnostic clinical risk prediction model calibration if the predictors are properly included in the model. Furthermore, using DAGs to identify Markov Blanket variables may be a useful, efficient strategy to select predictors in clinical risk prediction models if strong knowledge of the underlying causal structure exists or can be learned.

**Background**

In modern epidemiology, prediction modeling and causal inference are generally considered separate branches with unique sets of methods and aims. However, recently, the emerging field of "causal learning" or "causal discovery" has led to the introduction of prediction modelling and machine learning techniques as tools to generate structural causal models based on data-driven procedures[1]. Movement in the other direction has been less explored; namely, the application of causal inference principles and graph theory in clinical risk prediction modeling strategies.

Diagrams and graphs are intuitive, visual tools used to inform analytic methods to answer causal questions[2]. The increasing use of causal graphs and the need for automated procedures to assess causal effects given the combination of previous structural knowledge and new data led to the development of a compact, formal theory free of parametric assumptions to transparently model causal relationships[2]. Directed Acyclic Graphs (DAGs) are used to rigorously map all *a priori* assumptions surrounding a causal question of interest[2] and to graphically describe the underlying data generating process. In DAGs, each node represents a random variable, and directed causal paths are represented by arrows. The causal graph structure thus provides qualitative information about the conditional independencies of the variables of interest. DAGs are used as a tool in causal inference to illustrate potential sources of confounding and selection bias and ultimately identify suitable strategies to address them[3]. We assume the reader is familiar with DAGs; for those not yet familiar, several accessible introductions have been published elsewhere[2, 4].

The aim of this work is to investigate the potential benefits of using DAGs and causal thinking in clinical risk prediction problems. Specifically, we describe the use of causal knowledge in assessing transportability and selecting predictors for a clinical risk prediction model.

**Transportability and the principle of independent mechanisms**

A causal concept that could be useful in clinical risk prediction modeling is the *principle of independent mechanisms[1]*. This fundamental assumption was introduced to justify the inference of causal structure from observed data[1, 5] and was later suggested as a useful hypothesis to drive machine learning-based prediction approaches[6].

This principle of independent mechanisms states that the "causal generative process of a system's variables is composed of autonomous modules that do not inform or influence each other"[1]. This means that a causal process can be interpreted as a chain of independent mechanisms, in which each causal mechanism takes the state output from the previous mechanism as input and "feeds" the next mechanism with its own state output. Each causal mechanism on the chain can be conceptualized as a physical mechanism invariant to the input it receives[1]. The idea of the autonomy of the mechanisms is actually more intuitive than it seems. In fact, it is how we justify all clinical interventions: we assume that artificially changing one mechanism or its input will not affect any of the other mechanisms[1].

Let's consider two variables with an unconfounded causal relationship. For simplicity, we will call these two variables 'Cause' and 'Effect.' The joint probability distribution of these two variables $\mathbb{P}$(Cause,Effect) can be factorized in two ways[1, 6]:

$\mathbb{P}$(Cause,Effect) = $\mathbb{P}$(Effect|Cause)$\mathbb{P}$(Cause) = $\mathbb{P}$(Cause|Effect)$\mathbb{P}$(Effect)

The principle of independent mechanisms states that the marginal distribution of the variable Cause, $\mathbb{P}$(Cause), and the conditional distribution of the variable Effect on the variable Cause, $\mathbb{P}$(Effect|Cause), contain no information about each other[1, 6]. Indeed, $\mathbb{P}$(Effect|Cause) is the distribution of the variable Effect for each given value of the variable Cause. It represents the physical mechanism that transforms the input (Cause) into an output (Effect), while ℙ(Cause) represents the state of the input. Under the principle of independent mechanisms, ℙ(Cause) and ℙ(Effect|Cause) change independently of each other across different joint distributions[1].

This independence constraint in the first factorization induces a dependency between the conditional distribution of Cause on Effect, ℙ(Cause|Effect), and the marginal distribution of the Effect, ℙ(Effect), shown in the second mathematical factorization in the anticausal direction[1, 6]. Therefore, ℙ(Effect) and ℙ(Cause|Effect) often change in a dependent way across different joint distributions[1]. Since this concept of independence involves mechanisms rather than variables, it cannot be simply defined, tested, or quantified like the concept of statistical independence in probability theory[1].

In the next paragraphs, we will use two hypothetical, simplified clinical examples to illustrate the consequences of the *principle of independent mechanisms* in the context of diagnostic clinical risk prediction models.

**Example 1**

Say that we are interested in building a diagnostic clinical risk prediction model for the presence of Alzheimer's disease (Y=1), using the APOE ε4 allele status (X=1, presence; X=0, absence) as the sole predictor of the outcome in the general population of older persons. Y=0 indicates disease absence.

Since APOE ε4 is a known cause for Alzheimer's disease,[7] we could draw the DAG shown in Figure 1. Note that we are assuming a direct, unconfounded causal relationship (a strong assumption). By convention, each variable in the DAG is affected by a "noise" variable, which are assumed to be independent of other noise variables and modeled as random variables. These are usually not explicitly depicted because they are not of relevance to the causal relationship under study. However, it is worth noting that the noise variable affecting X determines the prevalence of the APOE ε4 allele, while the noise variable affecting Y contributes to the definition of the causal mechanism between the APOE ε4 allele status and Alzheimer's disease[6].

Assume we collect cross-sectional data about Alzheimer's disease and APOE ε4 allele status in a population A. Using this data, we can develop a simple diagnostic clinical risk prediction model using logistic regression to predict the presence of Alzheimer's disease. The regression equation would be:

$$\log_e(\Pr(Y=1|X=x)/\Pr(Y=0|X=x)) = \beta_0 + \beta_1 x$$

Using the logistic regression equation it's possible to estimate the four conditional probabilities $\Pr(Y=1|X=0)$, $\Pr(Y=1|X=1)$, $\Pr(Y=0|X=0)$, and $\Pr(Y=0|X=1)$, which define the conditional distribution $\mathbb{P}(Y|X)$.

We will assume for simplicity that the logistic regression is able to fully describe this conditional distribution, while the prevalence of the APOE ε4 allele ($\Pr(X=1)$) defines the marginal distribution $\mathbb{P}(X)$ of this predictor.

Next, say we want to use our newly developed risk prediction model as a diagnostic tool for Alzheimer's disease in another population B in which we know there is a different prevalence of the APOE ε4 allele. The new distribution of the predictor X in population B can be denoted as $\mathbb{P}^*(X)$.

According to the principle of independent mechanisms, the fact that the original distribution of X, $\mathbb{P}(X)$, has been changed to $\mathbb{P}^*(X)$ does not give any information on the mechanism $\mathbb{P}(Y|X)$ in population B.[1, 6] This is because X causes Y, and $\mathbb{P}(\text{Cause})$ is independent of $\mathbb{P}(\text{Effect}|\text{Cause})$.

If the underlying causal mechanism is not altered ($\mathbb{P}(Y|X)$ is the same in the two populations), the diagnostic clinical risk prediction model developed in population A will produce valid estimates also in population B. On the other hand, if the causal mechanism changed, knowing the predictor distribution $\mathbb{P}^*(X)$ does not give us any information about how the mechanism changed.[1, 6] In this case, the logistic regression model developed in population A for modeling $\mathbb{P}(Y|X)$ is still our best diagnostic tool candidate.[1, 6]

In this example, knowledge of the underlying causal structure suggests, that using the same diagnostic clinical risk prediction model in the new population is a reasonable choice.[1, 6]

**Example 2**

Say we are still interested in building a clinical diagnostic risk prediction model for the presence of Alzheimer's disease, but instead choose to use a different variable as the sole predictor, which indicates whether the concentration of tau protein in cerebrospinal fluid (CSF-tau) is above a predefined threshold. As before, Y=1 and Y=0 indicate presence and absence of Alzheimer's disease. K=1 indicates high tau protein concentration, and K=0 indicates low tau protein concentration.

It is known that high CSF-tau levels are associated with the presence of Alzheimer's disease. Specifically, as a consequence of the deposition of proteins in the brain that characterizes Alzheimer's disease, the concentration of tau protein is altered in the cerebrospinal fluid.[8] Therefore, the high level of tau protein in the cerebrospinal fluid can be interpreted as a consequence of Alzheimer's disease, leading to the DAG shown in Figure 2.

In this example, we identify Alzheimer's disease by its underlying pathological process instead of the definition based on diagnostic criteria. However, in the real world, direct effects are usually incorporated as part of the diagnostic criteria of the disease for practical clinical purposes. We further assume a direct effect of Y on K without confounding, even though we acknowledge direct effects of a disease are typically also caused by risk factors for the disease (introducing confounding in the Y→ K causal relationship depicted in Figure 2). These strong assumptions are needed to create a simplified, illustrative example.

As before, assume we have collected cross-sectional data about Alzheimer's disease and CSF-tau concentration in a new population C. Using population C data, we can develop another simple diagnostic clinical risk prediction model to predict Alzheimer's disease using logistic regression. The estimated regression equation would be:

$$\log_e(\Pr(Y=1|K=k)/\Pr(Y=0|K=k)) = \gamma_0 + \gamma_1 k$$

Assuming that logistic regression is suitable, its equation fully describes the underlying conditional distribution $\mathbb{P}(Y|K)$, while the prevalence of the high CSF-tau ($\Pr(K=1)$) defines the marginal distribution $\mathbb{P}(K)$ of the predictor.

Say that we now want to apply this clinical diagnostic risk prediction model developed in population C to detect the presence of Alzheimer's disease in a population D with a different prevalence of high CSF-tau concentration. However, we are now in an anticausal scenario in which we are trying to use the effect, CSF-tau concentration, to detect the cause, Alzheimer's disease. Therefore, $\mathbb{P}(Y|K)$ does *not* represent a causal mechanism and is *not* independent of $\mathbb{P}(K)$.

Since the marginal distribution of CSF-tau levels changes from $\mathbb{P}(K)$ in population C to $\mathbb{P}^*(K)$ in population D, a change in the conditional distribution, $\mathbb{P}(Y|K)$, is likely to occur because we are in an anticausal direction[1, 6]. The model developed in population C to describe $\mathbb{P}(Y|K)$ will not be well calibrated for use in the population D because the underlying conditional distribution of Y on K is different in the two populations. This would also hold if the causal mechanism that leads from Alzheimer's disease to the high CSF-tau concentration was the same in the two populations, as the equation describing the conditional distribution of Y on K is purely a mathematical artefact and does not describe the causal process.

No common causes of Y and K were included in this oversimplified Example 2, and we acknowledge that the transportability of the diagnostic clinical risk prediction model to different populations in similar anticausal scenarios would be higher if the predictor and the disease share one or more common cause(s). Still, through these simple examples, we challenge the generally accepted notion that a change in the distribution of the predictors does not affect diagnostic clinical risk prediction model calibration if the predictors are properly included in the model. As illustrated in Example 2, this may not hold true in anticausal scenarios, in which the predicted outcome is the disease and the predictor is a causal effect of the outcome[1, 6].

The idea that risk prediction models including the direct causes of an outcome of interest as predictors will be more transportable to different settings is also exploited in the causal learning "invariant causal prediction" method[1] and in the machine learning practice of "covariate shift"[1, 6]. In general, we think the field of diagnostic clinical risk prediction modeling could greatly benefit from the practice of incorporating knowledge of the underlying causal structure in modelling strategies. The integration of such information could provide insights into the transportability of a given diagnostic risk prediction model in different settings[6].

**Predictor selection and the Markov Blanket**

The first step in building a clinical risk prediction model is predictor selection. We focus on the main challenge of selecting the smallest possible subset of all available variables that provide enough information to predict the outcome of interest with good validity.

There are many reasons to limit the number of predictors used to build a risk prediction model: (i) to reduce problems due to the high number of variables in the model, thereby increasing performance, (ii) to reduce the costs, time and effort associated with data collection and storage, model development or training, (iii) to enable easier use of the model in different settings, and (iv) to increase the interpretability of the mechanisms behind the generation of the probability estimates[9, 10]. The last reason is particularly important in the context of clinical risk prediction models. Indeed, medical doctors are reluctant to use prediction models without a certain degree of interpretability[11], since the output probabilities are used to support clinical decisions about treatments and prevention strategies.

Intuitively, the predictor selection problem can be interpreted as how to choose the smallest subset of variables excluding all variables that do not provide additional information on the outcome of interest.

By operationalizing the lack of additional information using the notion of conditional independence,[12] the entire problem of predictor selection is analogous to identifying the so-called "Markov Blanket" of the outcome variable.

We define Y as the random variable for the outcome of interest and **X** as the set of all available candidate predictor variables of Y. We assume that **X** is a superset of the variables relevant to the causal processes in which Y is involved. The Markov Blanket of Y, MB(Y), is the minimal subset of **X**, conditioned on which, all other variables of **X** *not* included in MB(Y) are independent of Y[9, 10]:

$\forall$ V $\in$ **X** - MB(Y): Pr(Y|MB(Y),V) = Pr(Y|MB(Y)),

where **X** - MB(Y) denotes the set of variables which are contained in **X** but not in MB(Y). The concept of the Markov Blanket was first introduced by Pearl in 1988 in his work on Bayesian networks[13]. Years later, it was first used to identify the theoretical optimal set of variables for prediction tasks[12].

According to the definition above, given MB(Y), the other variables contained in **X** are independent of the outcome Y. This means that they do not provide any further information about Y, and all the information to predict the behavior of the outcome is already contained in the Markov Blanket MB(Y)[1, 14].

If the technique used to build the prediction model for Y can fully describe the underlying true probability Pr(Y|MB(Y)), and a model with fewer variables is preferred, then the variables included in the Markov Blanket of the outcome Y are the only variables needed for an optimal prediction[9]. Therefore, in an idealized regression setting, to fit the appropriate model, the predictor selection task consists of finding the Markov Blanket of the outcome variable[10]. This concept can be used to link variable selection in clinical risk prediction modeling to the underlying causal structure of the data[15].

Let's consider a DAG *G* and a set of variables S described by a joint distribution $\mathbb{P}_S$ with a density. The distribution $\mathbb{P}_S$, is said to be Markovian with respect to *G* if each variable is conditionally independent of its non-descendants (i.e. variables it does not affect), given its parents (i.e. its direct causes)[1, 10]. This Markov property creates a link between $\mathbb{P}_S$ and *G*, ensuring that all the conditional independencies entailed by the DAG are also present in the probability distribution[1, 16].

A further condition makes this link stronger; "faithfulness" states that the only conditional independencies to hold in the joint distribution $\mathbb{P}_S$ are the ones entailed in *G[15]*.

The previous intuition can be formalized; it has been demonstrated that if the joint distribution of the variables is faithful and Markovian with respect to the DAG, a predictor is strongly relevant[17] for predicting the outcome if and only if it is part of the Markov Blanket of the outcome[18]. Under these conditions, the Markov Blanket of the outcome is unique and has a particular constitution: it includes all parents of the outcome node, all of its children, and all parents of its children[1, 9, 10, 13].

As shown in Figure 3, these nodes "shield" the outcome variable Y from all the remaining variables in the DAG[14]. Therefore, the information contained in these nodes is sufficient to describe the outcome variable's status.

These results are appealing for researchers tasked with selecting predictors for clinical risk prediction modeling. According to a 2010 review, at least 8 different algorithms have been developed to identify the Markov Blanket for an outcome variable using data-driven procedures[10]. In the field of causal learning, algorithms that learn the entire causal structure[15] and the local causal structure[19] based on the identification of Markov Blankets have been developed.

Given this background, we argue that a strong knowledge of the underlying causal processes behind the data generation could substantially help to identify the best predictors to be included in a clinical risk prediction model.

As proof of concept, we conducted a series of simulations using R version 3.6.1. We simulated 100,000 datasets with 25 variables and 10,000 observations each. Each dataset was simulated according to a randomly generated DAG (using the randomDAG function in the dagitty R package). The DAG included 25 ordered nodes corresponding to 25 variables. Each node was given a probability of 0.1 of receiving a directed arrow from each of the individual previous nodes. One of the nodes was then randomly selected as the binary outcome of interest, all other 24 variables were assumed to be continuous. Any exogenous variables (i.e. variables without any parent nodes) were generated as normally distributed variables with a mean of 0 and variance of 1, or, if the outcome was exogenous, as a Bernoulli random variable with an event probability of 0.2.

When the outcome was an endogenous variable (i.e., with at least one parent node), each observation was drawn from a Bernoulli distribution with a defined probability parameter. This was set as the inverse-logit function evaluated at the linear combination of the outcome node's parent variables, with randomly drawn coefficients. Specifically, the coefficients (including the intercept) for each endogenous variable were drawn from a uniform distribution on (-1,1).

Similarly, the observations of the continuous endogenous variables were randomly drawn from a normal distribution with unit variance and with the mean equal to the linear combination of randomly drawn coefficients and the values of the node's parent variables. Here, the coefficients (including the intercept) for each endogenous variable were drawn from a uniform distribution on (-2,2). The choice of the regression coefficients was therefore not restricted in order to satisfy the faithfulness assumption by design.

For each of the 100,000 datasets, eight prediction tools were developed to predict the probability that the binary outcome equals 1:

(i) a logistic regression model including only variables in the Markov Blanket of the outcome as predictors,

(ii) a logistic regression model including all 24 variables as predictors,

(iii) a logistic regression model including any variable with a path leading to the outcome node (regardless of arrow direction on the path) as predictors,

(iv) a logistic regression model including only the outcome node's parent variables as predictors,

(v) a logistic lasso regression model inputting all 24 variables,

(vi) a logistic ridge regression model inputting all 24 variables,

(vii) a logistic elastic net regression model with mixing parameter alpha of 0.5 inputting all 24 variables, and

(viii) a random forest algorithm inputting all 24 variables.

Lasso, ridge, and elastic net models were computed using the glmnet function in the glmnet R package with default settings. The regularization parameter, lambda, that minimized the 10-fold cross-validated error based on the deviance for logistic regression with the cv.glmnet function (glmnet package) was selected. Random forests were built using the randomForest function in the randomForest R package with 1,000 trees and default settings.

For each dataset, the calibration of each prediction tool was measured using the Integrated Calibration Index[20] (ICI) based on 10-fold cross-validation. Lower ICI indicates better model calibration. The ICI estimation relies on a non-parametric regression between the outcome variable and the predicted risk estimated by the prediction tool. Therefore, if the non-parametric regression fails in one or more of the 10 cross-validation sets, it is not possible to compute the ICI. This happens if an intercept-only model or a model with variables' regression coefficients very close to 0 is evaluated. Summary performance metrics from the 100,000 simulated datasets are reported in Table 1.

In 37,281 of the simulated datasets, the outcome variable node did not have any parents, therefore it was not possible to assess the performance of logistic regression including only the outcome node's parent variables as predictors in these cases (Table 1). In 8,122 simulated datasets, the outcome variable node did not have any parents or children, therefore it was not possible to assess the performance of the Markov Blanket-based logistic model and the logistic regression including all the variables with a path to the outcome as predictors (Table 1).

When the Markov Blanket set was empty, both the lasso and elastic net regression models correctly shrunk all regression coefficients to zero or very close to zero 94% of the time, leading to an uncomputable ICI. Overall, the lasso regression selected exactly the Markov Blanket set of variables in at least one of the ten cross-validations in 15,515 (15.5%) simulated datasets. The percentage was higher when the Markov Blanket was empty (85.1%), included only one (47.6%) or only two (12.3%) variables. This finding supports the idea proposed by Li et al. that there is a strong link between the lasso regularization and selection algorithm and the identification of the Markov Blanket[21].

Overall, the average ICI of the Markov Blanket-based logistic model (0.01882) was lower compared with all other investigated prediction tools. This model also yielded the lowest average ICI (0.01956) when considering only those datasets in which all prediction tools had computable ICI values (Table 1). In head-to-head comparisons, the ICI of the various prediction tools were greater than or equal to the ICI of the Markov Blanket-based logistic model in the majority of the simulated datasets (range: 56.90% to 98.23%). Not only did the Markov Blanket-based logistic model show good performance in terms of calibration but also required considerably fewer input variables than the number of available variables.

These results empirically demonstrate equal or superior performance of the Markov Blanket-based logistic model, corroborating the theories presented earlier. We acknowledge that in real-world settings, it is unlikely to encounter ideal situations in which perfect knowledge of the underlying causal structure and all requisite variables are available and complete and non-linear relationships and interactions are absent. However, we believe our results provide an important contribution as a theoretical basis for using a DAG that summarizes *a priori* knowledge of the causal structure to identify predictors in a simple and structured way in an ideal setting.

**Conclusions**

Through a series of theoretical examples and simulation results, we have shown that strong knowledge of the underlying causal structure can be useful for understanding potential transportability and optimizing predictor selection for a given clinical risk prediction model. In the field of clinical risk prediction model development and application, we think that *a priori* causal information is often ignored or used intuitively without a structured framework. We are eager to see first applications of the framework we have outlined, further theoretical development, and scientific discussion of this concept.

**Acknowledgements:**

The authors wish to thank Dr. James M. Robins for insightful comments and suggestions, which helped improve this manuscript.

**Funding:**

JLR's research position was supported by a grant from the Else-Kröner-Fresenius Foundation (www.ekfs.de, GSO/EKFS-17, granted to TK). The funders had no role in study design, data collection and analysis, decision to publish, or preparation of the manuscript.

**Author Contributions:**

MP conceptualized the study. MP and SK designed and ran the simulations. MP and JR drafted the manuscript. TK and SK supervised the project. All authors critically reviewed the final version.

**Competing interests:**

We declare no conflicts of interest for the submitted work. JLR, MP and SK have nothing further to disclose. Outside of the submitted work, TK reports having contributed to an advisory board of CoLucid and a research project funded by Amgen, for which the Charité – Universitätsmedizin Berlin received an unrestricted compensation. He further reports having received honoraria from Lilly, Newsenselab, and Total for providing methodological advice, from Novartis and from Daiichi Sankyo for providing a lecture on neuroepidemiology and research methods, and from the BMJ for editorial services.

**Data Availability Statement:**

All data were simulated and the code is available upon request from the corresponding author (MP). The simulation conditions are described in detail in the manuscript along with software, package names and functions for full transparency.

**Figures and Tables**

**Fig 1.** Directed Acyclic Graph (DAG), Example 1

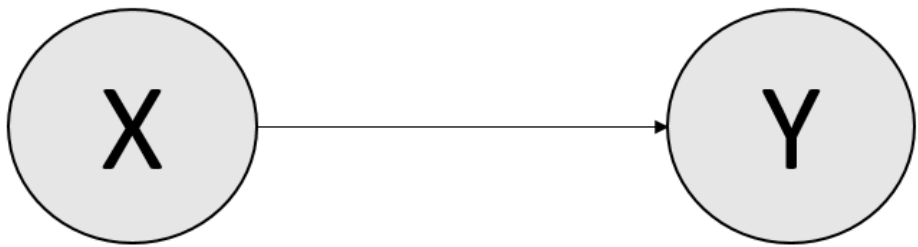

**Fig 2.** Directed Acyclic Graph (DAG), Example 2

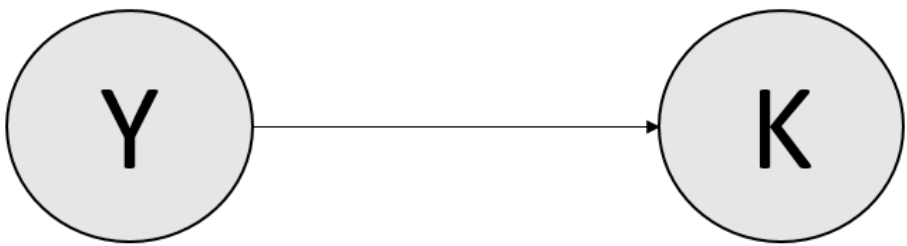

**Fig 3.** Example of the Markov Blanket (in black) of outcome Y in a simple Directed Acyclic Graph (DAG) with many nodes.

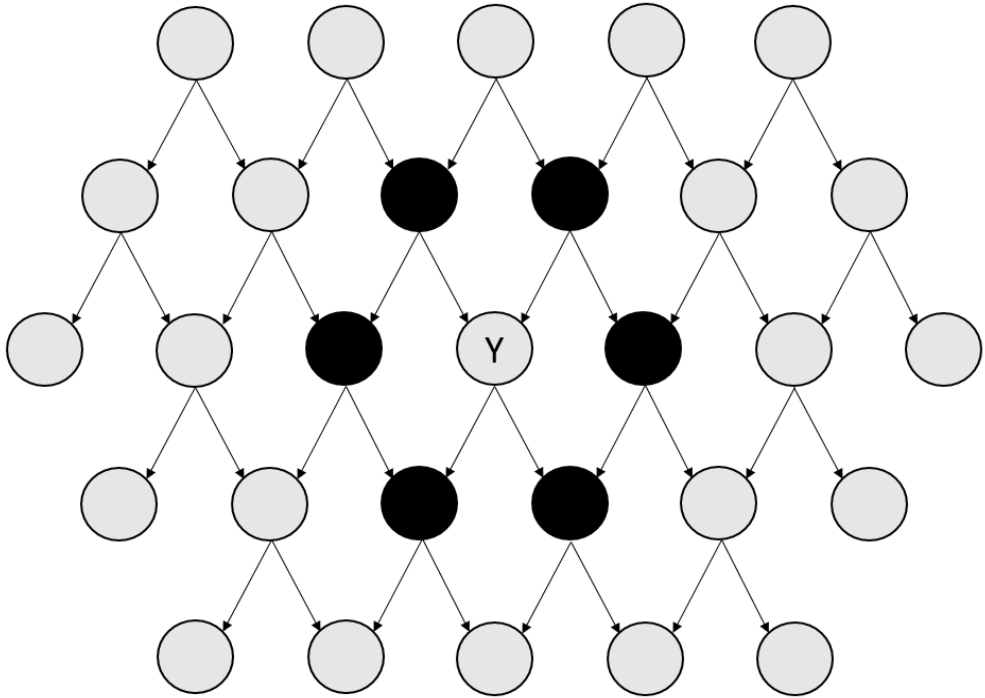

**Table 1.** Simulation Results: Prediction Tools' Performance Metrics

| | Logistic, Markov Blanket set (Nsim=100,000) | Logistic, all 24 variables (Nsim=100,000) | Logistic, any variables with a path to the outcome (Nsim=100,000) | Logistic, node's parent variables (Nsim=100,000) | Lasso, all 24 variables (Nsim=100,000) | Ridge, all 24 variables (Nsim=100,000) | Elastic net, all 24 variables (Nsim=100,000) | Random forest, all 24 variables (Nsim=100,000) |
|---|---|---|---|---|---|---|---|---|
| **FULL RESULTS: Including all simulated datasets** | | | | | | | | |
| **ICI** | | | | | | | | |
| N Missing | 8,122 | 0 | 8,122 | 37,281 | 8,743 | 0 | 8,733 | 1 |
| Mean (SD) | 0.01882 (0.00444) | 0.01965 (0.00495) | 0.01900 (0.00460) | 0.02215 (0.00422) | 0.01912 (0.00451) | 0.03809 (0.02065) | 0.01907 (0.00454) | 0.04138 (0.01778) |
| Median | 0.01857 | 0.01925 | 0.01866 | 0.02240 | 0.01889 | 0.02897 | 0.01881 | 0.03646 |
| Range | 0.00163 - 0.03702 | 0.00179 - 0.04535 | 0.00187 - 0.04094 | 0.00163 - 0.03803 | 0.00191 - 0.03726 | 0.00234 - 0.19527 | 0.00199 - 0.03758 | 0.00656 - 0.18998 |
| **Number of input variables** | | | | | | | | |
| N Missing | 0 | 0 | 0 | 0 | 0 | 0 | 0 | 0 |
| Mean (SD) | 4.0 (2.8) | 24.0 (0.0) | 18.9 (7.0) | 1.2 (1.3) | 24.0 (0.0) | 24.0 (0.0) | 24.0 (0.0) | 24.0 (0.0) |
| Median | 3 | 24 | 22 | 1 | 24 | 24 | 24 | 24 |
| Range | 0 - 20 | 24 - 24 | 0 - 24 | 0 - 8 | 24 - 24 | 24 - 24 | 24 - 24 | 24 - 24 |
| **Direct comparison: ICI of various methods compared to Markov Blanket-based logistic tool** | | | | | | | | |
| N Missing | 8,122 | 8,122 | 8,122 | 37,281 | 9,234 | 8,122 | 9,240 | 8,123 |
| < ICI logistic MB set, N (%) | | 39,317 (42.79%) | 39,597 (43.10%) | 4,933 (7.87%) | 26,359 (29.04%) | 8,878 (9.66%) | 31,071 (34.23%) | 1,630 (1.77%) |
| ≥ ICI logistic MB set, N (%) | | 52,561 (57.21%) | 52,281 (56.90%) | 57,786 (92.13%) | 64,407 (70.96%) | 83,000 (90.34%) | 59,689 (65.77%) | 90,247 (98.23%) |
| **COMPLETE CASE RESULTS: only including datasets for which ICI could be estimated for all tools** | | | | | | | | |
| **ICI** | | | | | | | | |
| N Missing | 37,872 | 37,872 | 37,872 | 37,872 | 37,872 | 37,872 | 37,872 | 37,872 |
| Mean (SD) | 0.01956 (0.00461) | 0.01975 (0.00476) | 0.01970 (0.00472) | 0.02212 (0.00421) | 0.01995 (0.00470) | 0.03896 (0.02185) | 0.01990 (0.00474) | 0.04059 (0.02010) |
| Median | 0.01953 | 0.01962 | 0.01961 | 0.02236 | 0.01993 | 0.02892 | 0.01987 | 0.03299 |

| Range | 0.00436 - 0.03702 | 0.00442 - 0.04016 | 0.00445 - 0.04094 | 0.00465 - 0.03803 | 0.00428 - 0.03726 | 0.00616 - 0.19527 | 0.00426 - 0.03758 | 0.00656 - 0.18998 |
|---|---|---|---|---|---|---|---|---|
| **Number of input variables** | | | | | | | | |
| N Missing | 37,872 | 37,872 | 37,872 | 37,872 | 37,872 | 37,872 | 37,872 | 37,872 |
| Mean (SD) | 4.1 (2.7) | 24.0 (0.0) | 20.8 (4.0) | 1.9 (1.1) | 24.0 (0.0) | 24.0 (0.0) | 24.0 (0.0) | 24.0 (0.0) |
| Median | 4 | 24 | 22 | 2 | 24 | 24 | 24 | 24 |
| Range | 1 - 19 | 24 - 24 | 1 - 24 | 1 - 8 | 24 - 24 | 24 - 24 | 24 - 24 | 24 - 24 |
| **Direct comparison: ICI of various methods compared to Markov Blanket-based logistic tool** | | | | | | | | |
| N Missing | 37,872 | 37,872 | 37,872 | 37,872 | 37,872 | 37,872 | 37,872 | 37,872 |
| < ICI logistic MB set, N (%) | | 26,986 (43.44%) | 27,321 (43.98%) | 4,920 (7.92%) | 16,818 (27.07%) | 6,530 (10.51%) | 20,025 (32.23%) | 1,618 (2.60%) |
| ≥ ICI logistic MB set, N (%) | | 35,142 (56.56%) | 34,807 (56.02%) | 57,208 (92.08%) | 45,310 (72.93%) | 55,598 (89.49%) | 42,103 (67.77%) | 60,510 (97.40%) |

Abbreviations: ICI, integrated calibration index; MB, Markov Blanket; Nsim, number of simulations; SD, standard deviation.

In a series of 100,000 simulated datasets, we obtained these results for ICI and number of input variables for the eight investigated prediction tools. Full results and complete case results, including only datasets for which ICI could be estimated for all tools are presented.